\documentclass[reprint,aip,jap,amsmath,amssymb,nofootinbib]{revtex4-1}
\usepackage{amsmath}
\usepackage{amsfonts}
\usepackage{amssymb}
\usepackage{graphics}
\usepackage{float}
\usepackage{graphicx}
\usepackage{epstopdf}
\usepackage[compact]{titlesec}
\usepackage{natbib}
\usepackage{threeparttable}
\usepackage[titletoc,title]{appendix}
\usepackage{lipsum}
\usepackage{textcomp}
\usepackage{subfigure}
\usepackage{color}
\usepackage[stable]{footmisc}
\bibpunct{[}{]}{,}{n}{}{}
\usepackage{appendix}

\newcommand{\etal}{\textit{et al}. }

\begin{document}
\title{The intrinsic inter-band optical conductivity of a C$_{2v}$ symmetric topological insulator}
\author{Parijat Sengupta}
%\email{psg@uic.edu}
\affiliation{Photonics Center
Boston University, Boston, MA, 02215.}

\begin{abstract}
In this work we analytically investigate the longitudinal optical conductivity of the C$_{2v}$ symmetric topological insulator. The conductivity expressions at $ T = 0 $ are derived using the Kubo formula and expressed as a function of the ratio of the Dresselhaus and Rashba parameters that characterize the low-energy Hamiltonian. We find that the longitudinal inter-band conductivity vanishes when Dresselhaus and Rashba parameters are equal in strength, also called the persistent spin helix state. The calculations are extended to obtain the frequency-dependent real and imaginary components of the optical conductivity for the topological Kondo insulator SmB$_{6}$ which exhibits C$_{2v}$ symmetric and anisotropic Dirac cones hosting topological states at $ \overline{X} $ point on the surface Brillouin zone. 
\end{abstract}
\maketitle

\vspace{0.25cm}
\section{Introduction}
\vspace{0.25cm}
Spectroscopy techniques serve as an important toolbox to probe the microscopic excitation of matter, the response of the electron ensemble to an external perturbation, and ground-state correlation functions that link to specific measurements. Spectroscopy measurements are broadly classified in to categories~\cite{coleman2015introduction} identified by the study of the spectrum of the microscopic variable within the matter; for example, angle-resolved photo-emission spectroscopy reveals the arrangement of electronic energy surface states while spin provides the basis for nuclear magnetic resonance, vital in determination of the structure of compounds. Similarly, the response of matter to light is routinely gauged using optical spectroscopy that utilizes the variation of the microscopic current density response to examine the the frequency-dependent electrical conductivity. The optical response of matter, in principle, is conveniently gauged from reflectivity data obtained from elipsometry techniques which provide information about the phase and amplitude of the reflected ray. In this regard we note, that governed by Maxwell's laws, the passage (and reflection) of light incident on matter and its eventual coupling to the intrinsic charge density has been studied for a wide class of materials spanning the entire gamut of classification. However, the emergence of topological states of matter modify Maxwell's laws through the introduction of a non-trivial $ \theta $ term~\cite{qi2008topological} that profoundly influences the final reflectivity pattern embodied in the optical conductivity behaviour. In this paper, we analytically derive expressions for inter-band optical conductivity arising from transitions between energy levels located on multiple bands in a topological insulator (TI) whose surface states are marked by the C$_{2v}$ symmetry. 

In principle, the surface dispersion of a topological insulator in its simplest form can be described by a Rashba-like linear Hamiltonian; however, the loss of intrinsic bulk symmetry (a crystal attribute characterized by the bulk inversion asymmetry parameter) in crystals necessitates the inclusion of an additional Dresselhaus-like Hamiltonian term.~\cite{winkler2003spin-orbit,vzutic2004spintronics} While the Dresselhaus contribution is usually a small effect and ignored in most calculations, we show that the overall ratio of the Rashba- and Dresselhaus-coupling coefficients can indeed have a significant role in modulating the electronic spectrum with important implications for light-matter interaction on the topological insulator surface. Recent developments in the fabrication of spin-based devices show that this ratio can be easily altered~\cite{yuan2013zeeman} by adjusting the Rashba coefficient which is a direct indicator of the structural inversion asymmetry (SIA). In fact, it has been experimentally confirmed that the Rashba parameter can be adjusted through an external gate bias~\cite{khomitsky2009electric,larionov2008electric} to values as large as 2$\times\,10^{-11} eV\,m $ in InAs-based heterostructures~\cite{grundler2000large}. Remarkably, for equal strength of the Rashba and Dresselhaus coupling coefficients (the persistent spin helix state), the inter-band optical conductivity vanishes.  

In this work, we utilize the Kubo formalism from linear response theory~\cite{jishi2013feynman} to establish a functional dependence between the ratio $ \left(\kappa\right) $   of the Dresselhaus $\left(\alpha_{D}\right)$ and Rashba $\left(\alpha_{R}\right)$ spin coupling coefficients $ \left(\kappa = \alpha_{D}/\alpha_{R}\right) $ and the longitudinal static components of charge and spin Hall conductivity. The examination of the conductivity in these systems yields useful information about the relative strength of the Rashba and Dresselhaus spin-orbit coupling coefficients; conversely, their relative strength provides key insight to the character of charge and spin conductance. This also allows us to note that while the sign of inter-band optical conductivity does not change as $ \kappa $ varies (the flow of charge to an external field is not regulated by spin), the overall spin texture and the sign of spin Hall conductivity (SHC) can be adjusted. Precisely, as $ \kappa $ takes on values greater than unity, the sign of SHC switches, a result also borne out by an identical transformation of the $ \pm\left(2n + 1)\right)\pi $ Berry phase of such a system. We further extend these calculations to obtain in the long wave length limit, the dynamic (frequency-dependent) real and imaginary parts of optical conductivity for the topological Kondo insulator~\cite{dzero2010topological,neupane2013surface} SmB$_{6}$. The topological surface states of SmB$_{6}$ possess anisotropic Dirac cones~\cite{1367-2630-17-2-023012} at the $\overline{X} $ point and characterized by C$_{2v}$ and time reversal symmetry. All calculations are performed at $ T = 0\,K$.

\vspace{0.25cm}
\section{Model Hamiltonians}
\vspace{0.25cm}
To describe the electronic structure of the material, we employ a two-band k.p model that describes surface states in proximity of the Dirac cone. In the low energy region, the linear Hamiltonian that describes the surface states is given by
\begin{equation}
H_{TI} = \alpha_{R}\left(\sigma_{y}k_{x} - \sigma_{x}k_{y}\right) + \alpha_{D}\left(\sigma_{x}k_{x} - \sigma_{y}k_{y}\right) + \Delta\sigma_{z},
\label{ham1}
\end{equation}
where $ \Delta $ is the symmetry-breaking potential that induces a finite gap between the surface bands while $ \alpha_{R} $ and $ \alpha_{D} $ are the Rashba- and Dresselhaus-like parameters, respectively. The dispersion relationship obtained by diagonalization of the Hamiltonian in Eq.~\ref{ham1} is $ \varepsilon\left(k\right) = \sqrt{\vert\,\beta\,\vert^{2}\,k^{2} + \Delta^{2}} $, where $ \vert \beta \vert^{2} =  \left(\alpha_{R}^{2} + \alpha_{D}^{2} + 2\alpha_{R}\,\alpha_{D}\,sin\,2\phi\right) $. For later use, we also write the analytic expressions for the $2\,\times\,1$ wave functions corresponding to the Hamiltonian in Eq.~\ref{ham1}
\begin{equation}
\Psi_{\pm} = \dfrac{1}{\sqrt{2}}\begin{pmatrix}
\lambda_{\pm}\exp\left(i\theta\right) \\
\pm\,\lambda_{\mp}
\end{pmatrix}; \lambda_{\pm} = \sqrt{1 \pm \dfrac{\Delta}{\sqrt{\Delta^{2}+\vert \beta \vert^{2}}}}.
\label{wf1}
\end{equation}
\noindent The polar angle in Eq.~\ref{wf1} is
\begin{equation}
\theta = tan^{-1}\dfrac{k_{x}+ \kappa\,k_{y}}{k_{y}+ \kappa\,k_{x}}
=  tan^{-1}\dfrac{\cos\phi+ \kappa\,\sin\phi}{\sin\phi+
\kappa\,\cos\phi}. 
\label{angpo}
\end{equation}
We have introduced the additional notation $ \kappa = \alpha_{D}/\alpha_{R} $ as the ratio of the Dresselhaus and Rashba coupling coefficients and set $k_{x} =
k\,cos\phi$ and $ k_{y} = k\,sin\phi $ in writing Eq.~\ref{angpo}. We also derive the corresponding velocity components $ v_{x,y} $ by evaluating the standard expression $ \hat{v}_{i} = (1/{i\hbar})\left[\hat{r},\mathcal{H}\right] $. The velocity components along \textit{x}- and \textit{y}-axes in operator notation are therefore $ \hat{v}_{x} = (1/\hbar)\left(\alpha_{D}\hat{\sigma}_{x} - \alpha_{R}\hat{\sigma}_{y}\right) $ and $ \hat{v}_{y} = (1/\hbar)\left(\alpha_{R}\hat{\sigma}_{x} - \alpha_{D}\hat{\sigma}_{y}\right) $, respectively. Finally, note that the k.p Hamiltonian written as an expansion of the states around the $ \overline{\Gamma} $ point is only accurate in low-energy regions in its vicinity. We can now using the model Hamiltonian (around the $ \overline{\Gamma} $ point) and the corresponding wave functions derive the longitudinal intra- and inter-band conductivity for TIs with C$_{2v}$ symmetry as a function of the Rashba and Dresselhaus coefficients.

\vspace{0.35cm}
\section{Intrinsic inter-band conductivity} 
\label{sec2}
\begin{figure}
\includegraphics[scale=0.42]{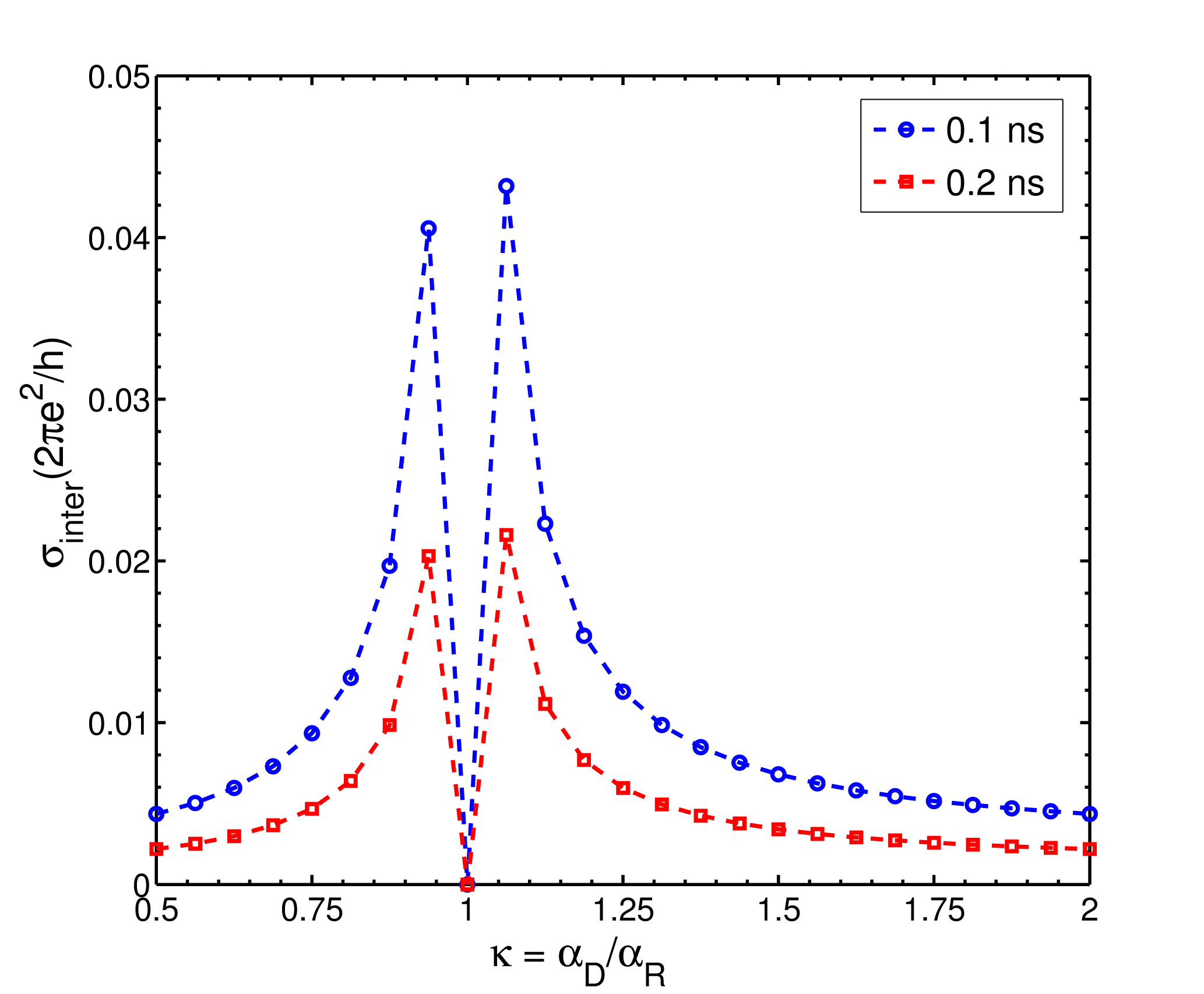}
\vspace{-10pt} \caption{The numerically calculated static inter-band optical conductivity for a range of $ \kappa = \alpha_{D}/\alpha_{R} $ and a pair of transit times $\left(\tau\right)$. At $ \kappa = 1 $, the inter-band optical conductivity vanishes. The static inter-band conductivity diminishes as the strength of the broadening parameter $ \eta = \hbar/\tau $ is reduced (for an increase in $ \tau $) eventually ceasing to exist for an infinite transit time. The Fermi level for this calculation was set to $ 5.0\, meV $.} 
\label{condco} 
\end{figure}

\vspace{0.35cm}
The conductivity calculations are carried out by a direct application of the Kubo formalism within the linear response theory. For a non-interacting sample, Kubo expression for conductivity is written as 
\begin{equation}
\sigma_{x,y} = -i\dfrac{\hbar\,e^{2}}{L^{2}}\sum_{n,n^{'}}\dfrac{f\left(\varepsilon_{n}\right)- f\left(\varepsilon_{n^{'}}\right)}{\varepsilon_{n} - \varepsilon_{n^{'}}}\dfrac{\langle\,n\vert\,v_{\alpha}\vert\,n^{'}\rangle \langle\,n^{'}\vert\,v_{\alpha}\vert\,n\rangle}{\varepsilon_{n} - \varepsilon_{n^{'}}+i\,\eta},
\label{kubof}
\end{equation}
where $ \vert\,n\rangle $ and $ \vert\,n^{'}\rangle $ are eigen functions of the defining Hamiltonian and $ \eta $ represents a finite broadening of the {eigen-states} resulting from surface imperfections and embedded impurities. We first take up the generic topological insulator with C$_{2v}$ symmetry and bulk inversion asymmetry described by Eq.~\ref{ham1}; these calculations with suitable amendment to the defining Hamiltonian are repeated for the topological Kondo insulator SmB$_{6}$. In deriving these expressions, we tacitly assume that the wave functions in presence of impurities and small external perturbations retain their original form given in Eq.~\ref{wf1} and the topological insulator sample area is $\mathcal{A} = L^{2}$. The inter-band conductivity expression (the intra-band expression is derived in Appendix~\ref{appA}) can be compactly written as
\begin{align}
&\sigma_{xx}^{inter} = -i\dfrac{\hbar\,e^{2}}{L^{2}}\sum_{n,n^{'}}\vert \mathcal{M}_{inter}\vert^{2}\biggl[\dfrac{f\left(\varepsilon_{c}\right)- f\left(\varepsilon_{v}\right)}{\left(\varepsilon_{c} - \varepsilon_{v}\right)\left(\varepsilon_{c} - \varepsilon_{v} + i\eta\right)}\biggr]. 
\label{intc1}
\end{align}
Note that the matrix element $ \mathcal{M}_{inter} $ in Eq.~\ref{intc1} for the inter-band case and a zero band gap (in Eq.~\ref{wf1}, $ \Delta = 0 $) is defined as
\begin{align}
\mathcal{M}_{inter} &= \langle \Psi_{\pm} \vert\,\hat{v_{x}}\vert\,\Psi_{\mp}\rangle = i\left(\alpha_{D}\sin\theta - \alpha_{R}\cos\theta\right),
\label{matxx}
\end{align}
where the valence and conduction band wave functions are $ \vert\,\Psi_{-}\rangle $ and $ \vert\,\Psi_{+}\rangle $, respectively and $ \hat{v}_{x} $ is the velocity operator along \textit{x}-axis. For a zero-temperature case $\left(T = 0\right)$, above the Fermi level, the conduction states are devoid of carriers while below the valence band is completely filled. The Fermi distribution functions $ f\left(\varepsilon_{c}\right) $ and $ f\left(\varepsilon_{v}\right) $ are therefore Heaviside step functions. We therefore set $ f\left(\varepsilon_{c}\right) $ and $ f\left(\varepsilon_{v}\right) $ to zero and unity, respectively. With this in mind, expanding and changing the sum in Eq.~\ref{intc1} to an integral, the final conductivity expression leads to:
\begin{align}
&\sigma_{xx}^{inter} = -i\dfrac{\hbar\,e^{2}}{4\pi^2}\int_{k_{f}}^{\infty}k\,dk\int_{0}^{2\pi}d\phi\left(\alpha_{D}f_{2}\left(\phi\right) - \alpha_{R}f_{1}\left(\phi\right)\right)^{2}\,\times \nonumber \\
& \biggl[\dfrac{\Theta\left(\varepsilon_{f} - \varepsilon_{c}\right) - \Theta\left(\varepsilon_{f} - \varepsilon_{v}\right)}{\left(2\varepsilon\right)\left(2\varepsilon + i\eta\right)}
+ \dfrac{\Theta\left(\varepsilon_{f} - \varepsilon_{v}\right) - \Theta\left(\varepsilon_{f} - \varepsilon_{c}\right)}{\left(-2\varepsilon\right)\left(-2\varepsilon + i\eta\right)}\biggr]. 
\label{intc2}
\end{align}
The functions $ f_{1}\left(\phi\right) $ and $ f_{2}\left(\phi\right) $ are $ \cos\left(\theta\right) $ and $ \sin\left(\theta\right) $, respectively expressed in terms of $ \phi $ using Eq.~\ref{angpo}. While writing Eq.~\ref{intc2}, for $ q \rightarrow 0 $ (in the long wavelength limit), we have set $ \varepsilon_{c} = -\varepsilon_{v} = \varepsilon $ and  $ \Theta\left(\cdot\right) $ represents the Heaviside step function. Simplifying the integrals in Eq.~\ref{intc2}, $ \sigma_{xx}^{inter} $ normalized to $ e^{2}/\hbar $ can be written as
\begin{align}
\begin{split}
\sigma_{xx}^{inter} &= \dfrac{\eta}{2\pi}\int_{0}^{2\pi}d\phi\dfrac{\left(\alpha_{D}f_{2}\left(\phi\right) - \alpha_{R}f_{1}\left(\phi\right)\right)^{2}}{\beta^{2}} \\
&\times \int_{k_{f}}^{\infty}\dfrac{1}{4\beta^{2}k^{2} + \eta^{2}}kdk.
\label{intc3}
\end{split}
\end{align}
Note that to change the variable of integration from $ k \rightarrow \epsilon $, we use the dispersion relation $ \epsilon = \beta\,k $. The Fermi energy is $ \varepsilon_{f} $ and $ k_{f} $ is the corresponding wave vector. The integral in Eq.~\ref{intc3} can be numerically evaluated to obtain $ \sigma_{xx} $. When expressed in terms of ratio of the Dresselhaus and Rashba coefficients, Eq.~\ref{intc3} is recast as
\begin{align}
\begin{split}
\sigma_{xx}^{inter}  &= \dfrac{1}{8\pi^{2}}\left(\pi/2 - tan^{-1}\dfrac{2\varepsilon_{f}}{\eta}\right)\\
&\times\,\int_{0}^{2\pi}\biggl[\dfrac{\left(\kappa^{2} - 1\right)sin\phi}{\left( \kappa^{2} + 2\kappa\,sin2\phi + 1\right)}\biggr]^{2}d\phi.
\label{fsx}
\end{split}
\end{align}
We immediately observe from Eq.~\ref{fsx} that for $ \kappa = 1 $, which defines a system with identical magnitude for the Rashba and Dresselhaus
coefficients - a condition known as the persistent spin helix (PSH) state with $ SU(2) $ symmetry occurs~\cite{bernevig2006exact,koralek2009emergence}- the longitudinal
static inter-band conductivity vanishes. The disappearance of the static inter-band conductivity can be simply explained by noting that the matrix element in the Kubo expression ceases to exist. Li \etal obtained an identical result in Ref.~\onlinecite{li2013vanishing} The inter-band optical conductivity, following a numerical integration of Eq.~\ref{fsx}, is shown in Fig.~\ref{condco} for two values of $ \eta $, the broadening parameter. 

Finally, we wish to point that the Hamiltonian in Eq.~\ref{ham1} is purely linear and, as a result, for the PSH state we obtained a vanishing longitudinal inter-band conductivity. However, there is always a cubic Dresselhaus contribution of the form $ \alpha_{D}^{'}\left(k_{x}k_{y}^{2}\hat{\sigma_{x}} - k_{y}k_{x}^{2}\hat{\sigma_{y}}\right) $ which leads to a finite conductivity. In this case, $ \alpha_{D}^{'} $ is the third order Dresselhaus coefficient and is the $SU(2)$ violating term. The vanishing inter-band optical conductivity also manifests as a zero inter-band absorption of light in the PSH state; a result which was also derived by the authors through an explicit calculation of the inter-band matrix element (the inter-band matrix element is zero in the PSH state) in connection to examining the circular dichroism $\left(\eta\right)$ pattern in TIs with C$_{2v}$ symmetry. Circular dichroism, which is the differential absorption of left- and right-circularly polarized light, for such a case has been derived in Ref.~\onlinecite{sengupta2016tunable}.
A note about dc conductivity is in order here: We have tacitly assumed $ T = 0 K $; however, the case of a finite temperature can also be easily handled by rewriting the Kubo expression in terms of the Matsubara Green’s function (or the imaginary time formalism). The conductivity expression derived in Eq.~\ref{fsx} is now a function of frequency and momentum $ \sigma\left(q,\omega\right)$, with $ q \rightarrow 0 $ in the long wavelength limit. The dc conductivity can be extracted from the general frequency-dependent expression by letting $ \omega \rightarrow 0 $.

Further, notice from Eq.~\ref{fsx} that the inter-band conductivity expression is always positive regardless of the relative strength of $ \kappa = \alpha_{D}/\alpha_{R} $, an observation easily reconcilable since charge conductivity does not depend on the orientation of the spin polarization brought about by the spin-orbit coupling Hamiltonian terms in Eq.~\ref{ham1}. However, there are quantities of interest, for instance, the Berry phase~\cite{sengupta2015influence} and the spin Hall conductivity~\cite{sengupta2016photo} that do exhibit a direct dependence on the strength of $ \kappa $. We examine the Berry phase in the following sub-section.

\vspace{0.25cm}
\subsection{The Berry phase}
\vspace{0.25cm}
The Berry phase is closely linked to the electron transport coefficients~\cite{sundaram1999wave}. We evaluate the Berry phase around the $\overline{\Gamma} $ point. The Berry phase \cite{grosso2014solid} in the closed $\overrightarrow{k} $- parameter space is defined as $ \gamma = \oint\,d\textbf{k}\cdot\,\langle\psi_{\pm},\theta
\vert\,i\dfrac{\partial}{\partial\,\textbf{k}}\vert\psi_{\pm},\theta
\rangle $.
Inserting the wave function from Eq.~\ref{wf1} in the Berry phase
$\left(\gamma\right)$ expression and evaluating $
\dfrac{\partial\theta}{\partial k_{\nu}} $ where $ \nu =
\left\lbrace x,y\right\rbrace $, one obtains
\begin{flalign}
\gamma = \int_{0}^{2\pi}\dfrac{\kappa^{2}-1}{\kappa^{2} + 2\kappa\,sin2\phi + 1}\,d\phi = \dfrac{\kappa^{2}-1}{\vert\,\kappa^{2}-1\,\vert}\pi,
\label{ber2}
\end{flalign}
where once more $ \kappa = \alpha_{D}/\alpha_{R} $. Let us now consider the two cases: 1) $ \vert\kappa\vert < 1 $ for which we have the Berry phase as $ -\pi $ using Eq.~\ref{ber2} while for 2) $ \vert\kappa\vert > 1 $, one obtains $ \gamma = \pi $. The switching of $ \kappa $ between the two aforesaid intervals, as we remarked before, does not change the optical conductivity; the significance of it, however, lies in its manifestation in the spin Hall conductivity. The zero-frequency spin Hall conductivity within the Kubo formalism is
\begin{equation}
\sigma_{SH} = -i\dfrac{\hbar\,e}{L^{2}}\sum_{n,n^{'}}\dfrac{f\left(\varepsilon_{n}\right)- f\left(\varepsilon_{n^{'}}\right)}{\varepsilon_{n} - \varepsilon_{n^{'}}}\dfrac{\langle\,n\vert\,j_{\alpha}^{z}\vert\,n^{'}\rangle \langle\,n^{'}\vert\,\hat{v}_{y}\vert\,n\rangle}{\varepsilon_{n} - \varepsilon_{n^{'}}+i\xi},
\label{kubosh}
\end{equation}
where $ \vert\,n\rangle $ and $ \vert\,n^{'}\rangle $ are eigen functions of the Hamiltonian given in Eq.~\ref{ham1}. The Kubo expression in Eq.~\ref{kubosh} when evaluated for $ \alpha = x $ yields the spin Hall conductivity. Note that the spin current operator~\cite{murakami2006intrinsic} is defined as $ j_{x}^{z} =
\dfrac{\hbar}{4}\left\lbrace \hat{v}_{x},\hat{\sigma}_{z}\right\rbrace $. In this definition of the spin current, the electron velocity is directed along the \textit{y}-axis due to an aligned external electric field and an out-of-plane $ \hat{z}-$polarized spin current flows along the perpendicular \textit{x}-axis. The spin Hall effect (SHE) refers to a transverse spin current induced by an external electric field in absence of a magnetic field. In this case, the SHE is produced by the intrinsic spin-orbit coupling as opposed to the extrinsic SHE driven by spin-orbit scattering impurities. The SHE leading to accumulation of spin in a preferred direction, as is easy to understand, depends on the strength of the two contributions to the Hamiltonian (Eq.~\ref{ham1}) which transform the spin components differently. It is therefore reasonable to believe that the path of spin accumulation in an intrinsic SHE setup guided by the spin-orbit coupling can be formulated in terms of the Berry phase. In the Kubo expression for spin conductivity (Eq.~\ref{kubosh}), inserting the desired eigen values and eigen functions and following
exactly the same set of steps carried out for optical conductivity, we arrive at the following expression for the spin Hall conductivity\footnote[3]{The spin current operator definition does not hold along the \textit{x} and \textit{y}-axes since the in-plane spin vectors are mixed as evident from Eq.~\ref{ham1}. The spin Hall conductivity in this case must be computed by evaluating the quantity $ \sigma_{xy}^{\varsigma,\uparrow} - \sigma_{xy}^{\varsigma\downarrow}$, where $ \varsigma = x,y $.}
\begin{equation}
\sigma_{xy}^{z} = \dfrac{e}{8\pi}\dfrac{1 - \kappa^{2}}{\vert\,1 - \kappa^{2}\,\vert}.
\label{sigmash}
\end{equation} 

In deriving the above expression, we added a particle-hole asymmetric quadratic term $ \hat{p}^{2}/2m $ to the Hamiltonian, following which the spin current operator evaluates to $ j_{\alpha}^{z} = \left(\hbar^{2}k_{\alpha}/2m\right)\sigma_{z} $, where $ \alpha =
\left\lbrace x,y\right\rbrace $. The connection between the geometric Berry phase and the spin Hall and diagonal conductivity can be easily seen by rewriting Eq.~\ref{sigmash} as $ \sigma_{xy}^{z} = \pm\,\left(e/8\pi^{2}\right)\gamma $. As the contribution of the two terms in the Hamiltonian reflected in the ratio $ \kappa $ toggles between the two intervals, $ \vert\kappa\vert < 1 $ and $ \vert\kappa\vert > 1 $, the spin Hall conductivity switches sign identically to the Berry phase. Furthermore, for the two limiting cases, $ \alpha_{R} = 0 $, when the system possesses structural inversion symmetry and a finite bulk inversion asymmetry, $\left(\alpha_{D} \neq 0 \right) $, and rewriting Eq.~\ref{sigmash} as $ \sigma_{xy}^{z} = e/8\,\pi\left(\alpha_{D}^{2} - \alpha_{R}^{2}\right)/\vert \alpha_{D}^{2} - \alpha_{R}^{2} \vert $, yields the universal spin Hall conductance as $ e/8\,\pi $. Conversely, for $ \alpha_{R} \neq 0 $ and $ \alpha_{D} = 0 $, the spin Hall conductance retains the same magnitude but switches sign. This result connecting the Berry phase to spin Hall conductivity was first obtained by S. Shen in Ref.~\onlinecite{shen2004spin}.

We again underscore the case of $\kappa = 1 $ which is the condition for PSH and recognize that spins are aligned parallel~\cite{schliemann2003nonballistic} and there is no ``effective" spin-orbit coupling that bends the trajectory. However, as we stated before, in the case of vanishing inter-band conductivity, non-zero higher-order Dresselhaus terms could lead to an inexact cancellation of the Rashba and Dresselhaus linear spin-orbit Hamiltonians. 

\vspace{0.25cm}
\section{Application to SmB$_{6}$} 
\label{sec2b}
\begin{figure}
\includegraphics[scale=0.25]{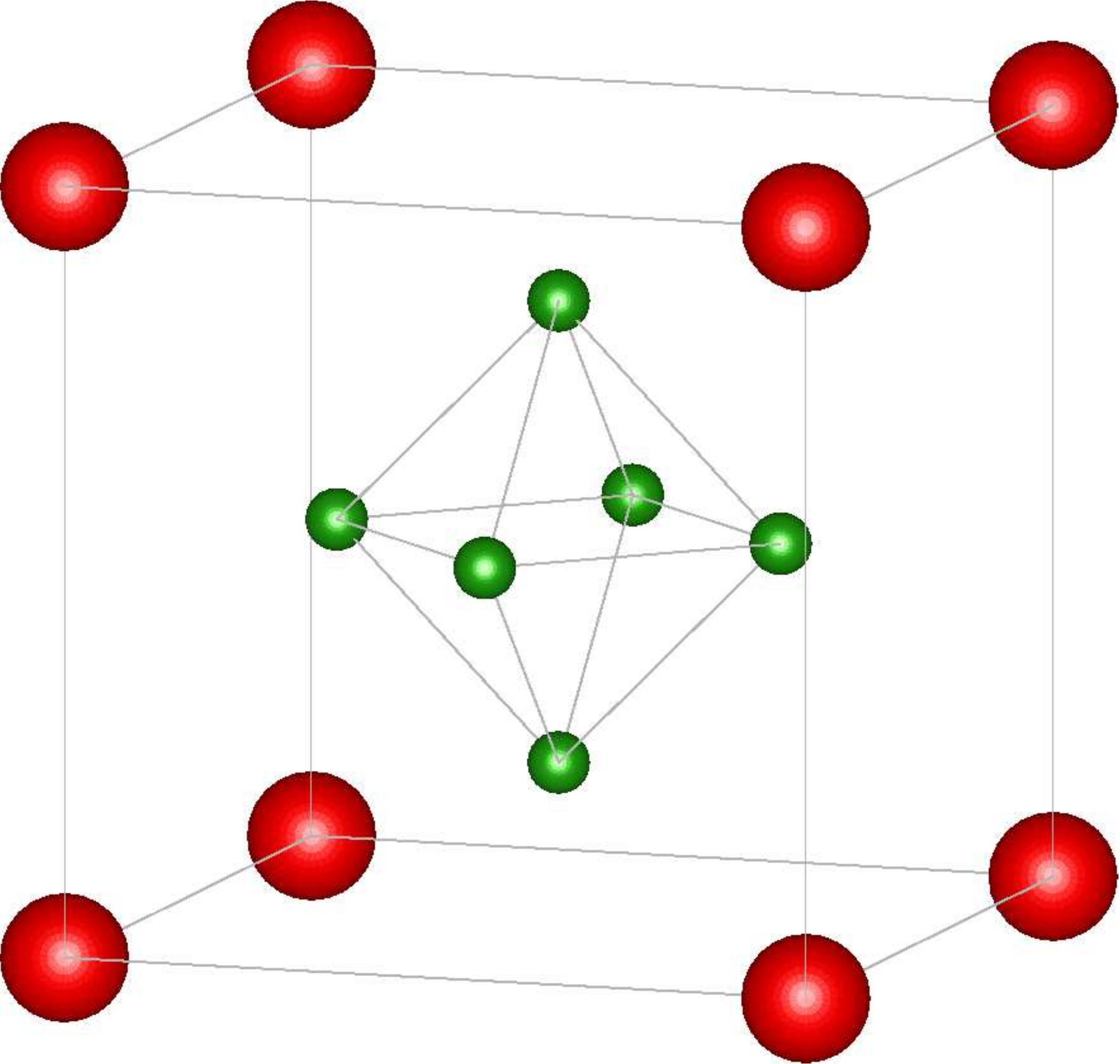}
\caption{The bulk unit cell for SmB$_{6}$ which has a CsCl-type crystal structure. The samarium (red) and boron (green) atoms arranged at the vertices of an octahedron are located at the corner and centre of a cubic lattice respectively. The structure visualization was done with the VESTA~\cite{momma2008vesta} software.}
\label{buc}
\end{figure}

\vspace{0.25cm}
A noteworthy instance of a material whose surface states have non-trivial topology with C$_{2v}$ symmetry is the topological Kondo insulator SmB$_{6}$ (see Fig.~\ref{buc} for the unit cell structure). Briefly, Kondo insulators which are marked by resistivity that has a minimum at a low temperature but increases as the temperature is lowered are highly electron-correlated systems that can exhibit the $ Z_{2} $ topological insulator behaviour. Experimental demonstrations of Kondo insulators with topologically non-trivial states have been carried out~\cite{wolgast2013low} with SmB$_{6}$ confirming their robust spin-polarized~\cite{kim2013surface,xu2014direct} surface states. SmB$_{6}$, however, unlike other Kondo insulators has a resistivity which has a minimum at room temperature $\left(T\right)$ and a plateau-like profile~\cite{allen1979large} for $\left(T \leqslant 5K \right)$. The low temperature constant resistivity is attributed to topologically protected surface states within the Kondo band gap (approximately 17.7 $ \mathrm{meV} $) and form three Fermi surfaces. Of these three Fermi surfaces, angle-resolved photoemission spectroscopy (ARPES) reveals that they are centred at the $ \overline{\Gamma} $ and doubly at the $ \overline{X} $ points of the surface Brillouin zone.

\begin{figure}
\centering
\includegraphics[scale=0.45]{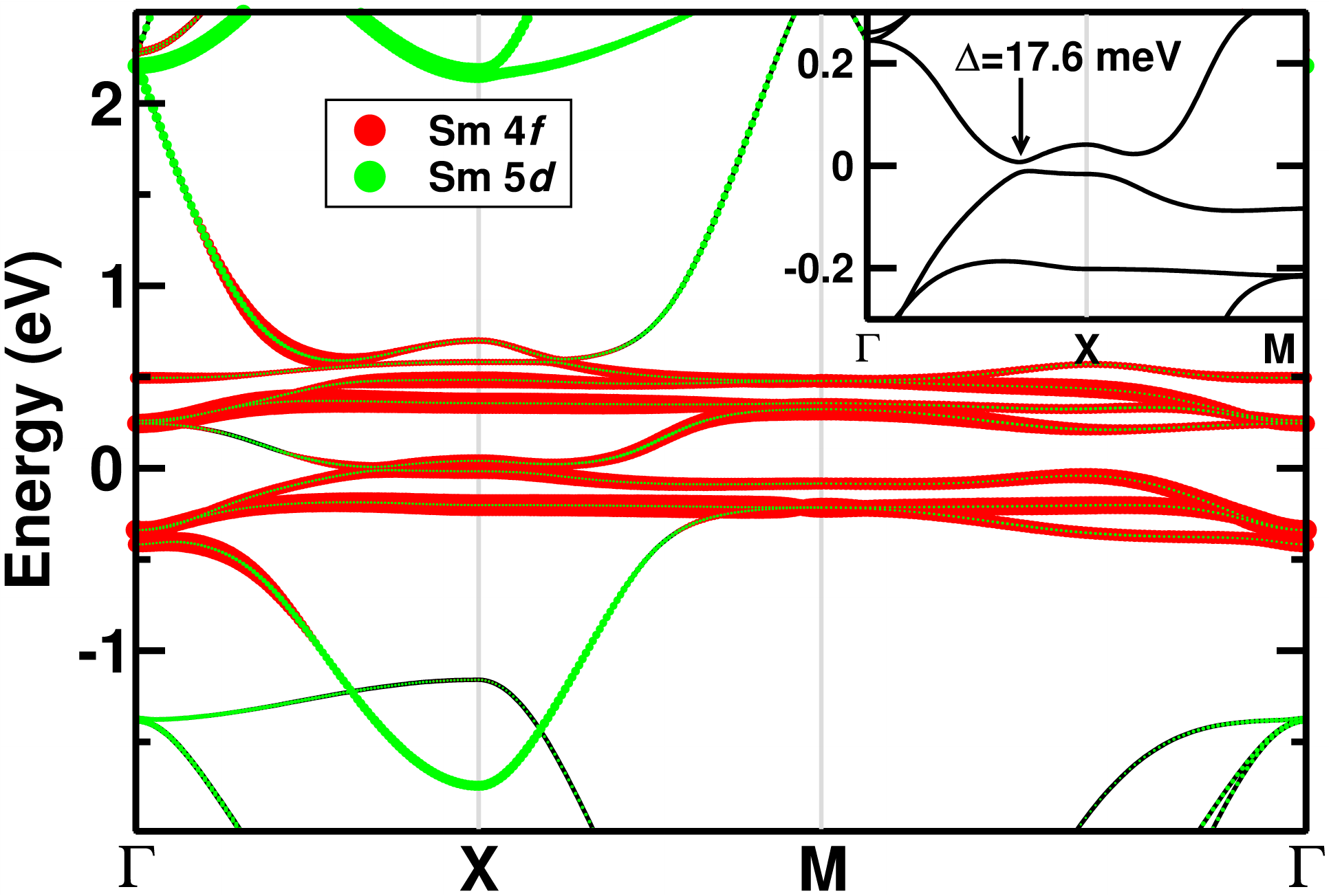}
\caption{The bulk band structure of SmB$_{6}$ with CsCl-type crystal structure calculated from first-principles using the VASP software. The inversion in the bulk band structure happens at the $ X $ point which upon projection to the surface manifests as three topologically protected states.}
\label{bulkbb}
\end{figure}

\begin{figure}
\includegraphics[scale=0.378]{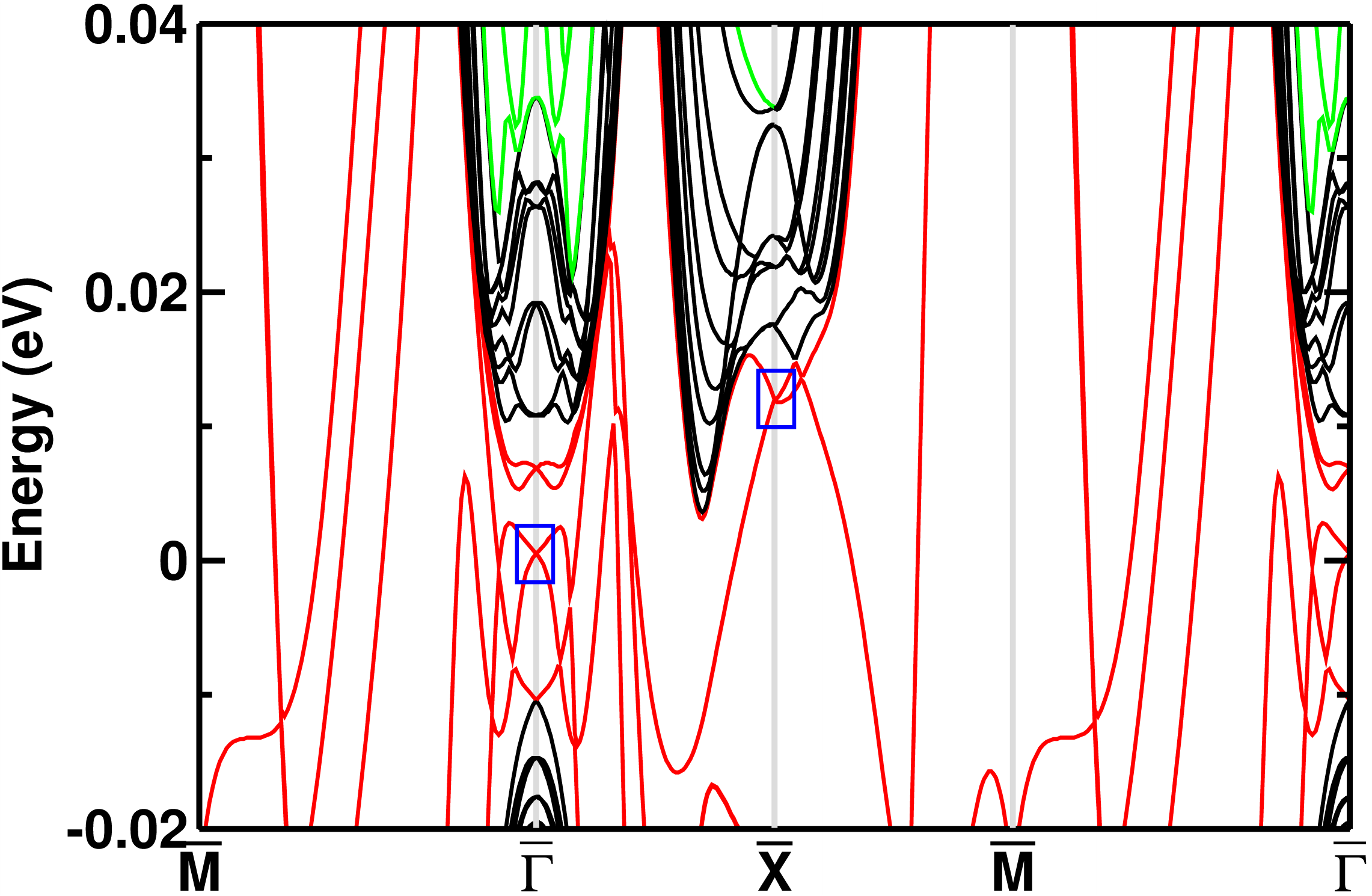}
\caption{The first-principles (with VASP) calculated dispersion of an SmB$_{6}$ slab of thickness $ 6.612\,nm $ is plotted with an inverted bulk character at $ X $. The inversion at bulk $ X $ is $\approx 17.7$ \, meV (see inset of Fig.~\ref{bulkbb}). The two topological surface states for the slab structure at $ \overline{\Gamma} $ and $ \overline{X} $ are boxed in blue. The slab configuration used in the VASP software package had 17 samarium atoms and 16 boron layers. The significance of the different colors on the plot is explained in the text.}
\label{bulksurf}
\end{figure}

The genesis of the topologically protected surface states at $ \overline{\Gamma} $ and $ \overline{X} $ lies in the band inversion~\cite{lu2013correlated} that occurs between the $ 4f $ and $ 5d $ orbitals of samarium at $ X $ points (see Fig.~\ref{bulkbb}) in the bulk Brillouin zone (BZ). These inverted $ X $ points in the bulk BZ when projected on the surface BZ of a $ \left[001\right] $ grown SmB$_{6}$ crystal gives rise to the Dirac-like surface states (see Fig.~\ref{bulksurf}) with a helical spin texture, the defining hallmark of topological insulators. A similar set of calculations was reported in Ref.~\onlinecite{chang2015two} where Tay-Rong Chang \etal computed the electronic structure of SmB$_{6}$ using the GGA and GGA + \textit{U} schemes; remarkably, they observed little change in the overall character of bands as the electrostatic Coulomb energy $ U $ was incremented from zero to a large value of $ 8.0\,eV $. They concluded by noting this functional non-dependence on the Coulomb energy as a sufficient proof of the band topology and the Kondo insulator attribute of SmB$_{6}$.
The bulk and slab band structures were obtained using the VASP code~\cite{kresse1996efficient,kresse1999ultrasoft} within Perdew-Burke-Ernzerhof~\cite{perdew1996generalized} exchange-correlation functionals. The spin orbit coupling was self-consistently included in our calculation. Additionally, the plane wave energy cutoff value was set to  $ 320\,eV $ and the Brillouin zone was sampled with a $ 12\times12\times12$ $\Gamma$ centered $k$-point mesh. We set the lattice constant~\cite{funahashi2010x} to $ a_{lc} = 4.1327\AA $ and the number of electrons with up and down spins were identical throughout the calculation. \textcolor{red}{Note that the dispersion of the slab in Fig.~\ref{bulksurf} uses three colors to indicate the origin of the bands: The red colored bands arise because of a dominant surface contribution while the black bands denote the bulk dispersion. The green bands are an admixture of surface and bulk dispersion. It is, however, pertinent to remember that only those surface bands that connect the conduction and valence bands (a closing of the band gap) are topological in nature; in this case they occur at the $ \overline{X} $ and $ \overline{\Gamma} $ points of the surface Brillouin zone and marked by blue rectangular contours. To obtain the surface contribution, we selected atoms that lie at the interface of the slab and vacuum.}

\begin{figure}[!b]
\centering
\includegraphics[scale=0.7]{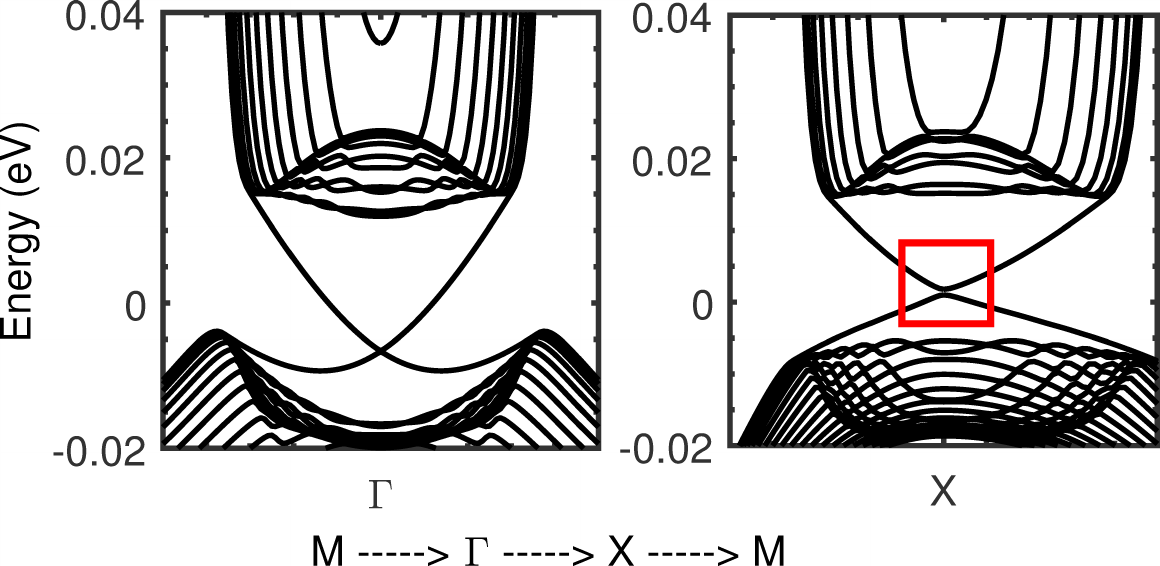}
\caption{\textcolor{red}{The 8-band \textit{k.p} calculated dispersion of an SmB$_{6}$ slab of thickness $ 7.0\,nm $ is plotted in the vicinity of $ \overline{\Gamma} $ and the $ \overline{X} $ point (right panel). The red boxed area in the right panel is the region of interest and we compute the inter-band conductivity by evaluation of the momentum matrix element (and their insertion in the Kubo expression) between states lying on the surface conduction and valence bands. The anisotropy of the Dirac cone around the $ \overline{X} $ point is clearly noticeable in contrast to the the isotropic Dirac cone centred at $ \overline{\Gamma} $. The Fermi level is aligned to top of the surface valence band.}}
\label{kpdisp}
\end{figure}

\textcolor{red}{For greater clarity, and to clearly identify the topological surface bands from the multiple states in a first-principles calculation, a simplified eight-band \textit{k.p} Hamiltonian that selectively describes the dispersion around the $ X $ point (derived from the theory of invariants in Ref.~\onlinecite{yu2015model}) is of utility in our context here. The bulk Hamiltonian is then adapted for a slab by carrying out the standard transformation $ k_{i} = -\partial/\partial k_{i} $; for our case, the dispersion of the slab around the $ \overline{\Gamma} $ and $ \overline{X} $ are obtained by making the simple replacement $ k_{z} = -i\partial/\partial z $ and  $ k_{x} = -i\partial/\partial x $ in the \textit{k.p} Hamiltonian. Note that the set of $k $-vectors for each case, $ \left(k_{x}, k_{y}\right) $, around the $ \overline{\Gamma} $ and $ \left(k_{y}, k_{z}\right) $ for $ \overline{X} $ continue to be good quantum numbers. Discretizing the derivative operator on a finite difference grid, we arrive at an effective slab Hamiltonian which can be diagonalized to obtain the dispersion as shown in Fig.~\ref{kpdisp}. For numerical details about discretization and other steps to construct the slab Hamiltonian, the reader is referred to Ref.~\onlinecite{sengupta2016numerical}.}
 
The anisotropic character of the Dirac cones around the $ \overline{X} $ point can be further reduced to an effective minimal \textit{k.p} surface Hamiltonian using symmetry arguments derived in Ref.~\onlinecite{yu2015model}. The band parameters for this model are obtained through a fitting procedure by a direct comparison with first principles calculation (Fig.~\ref{bulksurf}). The Hamiltonian at $ \overline{X} $ has a C$_{2v}$ symmetry which we reproduce in Eq.~\ref{tkix} and use as a starting point for further conductivity calculations with the topological Kondo insulator SmB$_{6}$.
\begin{equation}
H_{X}^{SS} = \epsilon_{1} + a_{0}k^{2} + \left[i\left(a_{1}k_{+} + a_{2}k_{-}\right)\sigma_{+} + h.c\right].
\label{tkix}
\end{equation}   
In Eq.~\ref{tkix}, the constants~\cite{yu2015model} (in units of $ a_{lc}*eV\AA $) $ a_{0}, a_{1}, a_{2} $ are 0.011276, 0.003059, and -0.02322, respectively. The other terms are defined as $ k_{\pm} = k_{x} \pm ik_{y}  $, $ \hat{\sigma}_{\pm} = \hat{\sigma}_{x} \pm i\hat{\sigma}_{y} $, and $ k^{2} = k_{x}^{2} + k_{y}^{2} $. The Pauli spin matrices are $ \sigma_{x},\sigma_{y} $, and $ \sigma_{z} $. Further, for low energy states the quadratic term in Eq.~\ref{tkix} can be dropped to yield a linear Dirac-like equation. Expanding, we obtain
\begin{equation}
H_{X}^{SS} = \left(a_{1} + a_{2}\right)k_{x}\sigma_{x} - \left(a_{1} - a_{2}\right)k_{y}\sigma_{y}.
\label{tkixs}
\end{equation}
Note that this represents an anisotropic (tilted) Dirac crossing at the $ \overline{X} $ point with unequal and \textit{x}- and \textit{y}-axes directed Fermi velocities.

To arrive at optical conductivity expressions for SmB$_{6}$ (Section~\ref{sec4a}), we begin (using the Hamiltonian in Eq.~\ref{tkixs}) by writing the wave functions for energy states in the vicinity of the anisotropic Dirac crossing. The conduction and valence state wave functions are
\begin{subequations}
\begin{equation}
\Psi_{\pm} = \dfrac{1}{\sqrt{2}}\begin{pmatrix}
\chi_{\pm}\exp\left(i\theta\right) \\
\pm\chi_{\mp}
\end{pmatrix},
\label{tkiwf1}
\end{equation}
where $ \chi_{\pm} $ is
\begin{equation}
\chi_{\pm} = \sqrt{1 \pm \dfrac{\Delta}{\sqrt{\Delta^{2} + \left(Ak_{x}^{2} + Bk_{y}^{2}\right)}}}.
\label{tkiwf2}
\end{equation}
\end{subequations}
The conduction (+) and valence (-) eigen states are defined as
\begin{equation}
\varepsilon_{\pm} = \pm\sqrt{\Delta^{2} + \left(A^{2}k_{x}^{2} + B^{2}k_{y}^{2}\right)}.
\label{tkiegs}
\end{equation} 
The conduction state wave function $ \Psi_{c} = \Psi_{+} $ and the corresponding valence state wave function is identified as $ \Psi_{v} = \Psi_{-} $. For brevity, in Eq.~\ref{tkiwf2}, $ A = a_{1} + a_{2} $ and $ B = a_{1} - a_{2} $. Note that for the sake of completeness a gap opening term of the form $ \Delta\sigma_{z} $ appears in the Hamiltonian (Eq.~\ref{tkixs}).

\vspace{0.25cm}
\subsection{Conductivity of the $\overline{X}$ point}
\label{sec4a}
\vspace{0.25cm}
The conductivity calculations again begin from the Kubo expression in Eq.~\ref{kubof}; however, to obtain a frequency dependence of the inter-band conductivity we must rework some of the expressions derived heretofore. The frequency-dependent Kubo expression for the inter-band conductivity is
\begin{flalign}
\sigma_{\alpha\beta}^{inter} = -i\dfrac{\hbar\,e^{2}}{L^{2}}\sum_{n,n^{'}}\dfrac{f\left(\varepsilon_{n}\right)- f\left(\varepsilon_{n^{'}}\right)}{\varepsilon_{n} - \varepsilon_{n^{'}}}\dfrac{\langle\,n\vert\,\hat{v}_{\alpha}\vert\,n^{'}\rangle \langle\,n^{'}\vert\,\hat{v}_{\beta}\vert\,n\rangle}{\hbar\omega + \varepsilon_{n} - \varepsilon_{n^{'}}+i\eta},
\label{kubofw}
\end{flalign}
where the symbols have meaning identical to Eq.~\ref{kubof}. For a less cumbersome notation, the superscript `inter' will be dropped from now. Further, the \textit{ket} vector $ \vert\,n\rangle\left(\vert\,n^{'}\rangle\right) $ denotes the conduction (valence) state wave function $ \vert\,\Psi_{c}\rangle \left(\vert\,\Psi_{v}\rangle\right) $. The corresponding conduction (valence) eigen state is $\varepsilon_{n = c} \left(\varepsilon_{n^{'}= v}\right) $. For longitudinal optical conductivity along the \textit{x}-axis, the velocity operators $ \hat{v}_{\alpha} $ and $ \hat{v}_{\beta} $ are identical and equal to $ \hat{v}_{x} = \left( A/\hbar\right)\hat{\sigma}_{x} $. Inserting the velocity operator in the Kubo expression, the matrix elements can be straightforwardly computed 
\begin{flalign}
\langle\,\Psi_{c}\vert\,\hat{v}_{x}\vert\,\Psi_{v}\rangle  = -\dfrac{A}{\hbar}\left(i\sin\theta + \gamma\cos\theta\right),
\label{matx}
\end{flalign}
where $ \gamma = \Delta/\sqrt{A^{2}k_{x}^{2} + B^{2}k_{y}^{2}} $. The Fermi distribution functions in Eq.~\ref{kubofw}, as before, are set under the tacit assumption that the Fermi energy is positioned at the top of the valence band; this effectively ensures that $ f\left(\varepsilon_{c}\right) = 0 $ and $ f\left(\varepsilon_{v}\right) = 1 $. Putting all of them together and expanding the real and imaginary part of the conductivity (in units of $ e^{2}/\hbar $) yields 
\begin{align}
\sigma_{xx}^{R} &= \eta\dfrac{A^{2}}{8\pi^{2}}\biggl[\int_{0}^{k_{c}}dk\int_{0}^{2\pi} d\theta\dfrac{\left(\sin^{2}\theta + \gamma^{2}\cos^{2}\theta\right)\Omega}{\left(\hbar\omega - 2\varepsilon\right)^{2} + \eta^{2}} + \nonumber \\
& \int_{0}^{k_{c}}dk\int_{0}^{2\pi} d\theta\dfrac{\left(\sin^{2}\theta + \gamma^{2}\cos^{2}\theta\right)\Omega}{\left(\hbar\omega + 2\varepsilon\right)^{2} + \eta^{2}}\biggr],
\label{tkirx}
\end{align}
where $ \Omega \approx \dfrac{1}{\sqrt{A^{2}\cos^{2}\theta + B^{2}\sin^{2}\theta}} $ and the upper limit of the \textit{k}-space integral, $ k_{c} $, corresponds to the momentum vector for a given energy cut-off. The approximation to $ \Omega $ is reasonably accurate for small values of $ \Delta $, the band gap opening. We set $ k_{c} = 0.12\,\AA^{-1} $ for a numerical evaluation of all conductivity expressions in this work. The imaginary part of the longitudinal optical conductivity is likewise,
\begin{align}
\sigma_{xx}^{Im} &= \dfrac{A^{2}}{8\pi^{2}}\biggl[\int_{0}^{k_{c}}dk\int_{0}^{2\pi} d\theta\dfrac{\left(\sin^{2}\theta + \gamma^{2}\cos^{2}\theta\right)\Omega^{'}_{-}}{\left(\hbar\omega - 2\varepsilon\right)^{2} + \eta^{2}} + \nonumber \\
& \int_{0}^{k_{c}}dk\int_{0}^{2\pi} d\theta\dfrac{\left(\sin^{2}\theta + \gamma^{2}\cos^{2}\theta\right)\Omega^{'}_{+}}{\left(\hbar\omega + 2\varepsilon\right)^{2} + \eta^{2}}\biggr],
\label{tkiimx}
\end{align}
where $ \Omega^{'}_{\pm} \approx \dfrac{\hbar\omega \pm 2\varepsilon}{\sqrt{A^{2}\cos^{2}\theta + B^{2}\sin^{2}\theta}} $.

\begin{figure}[!t]
\centering
\includegraphics[scale=0.7]{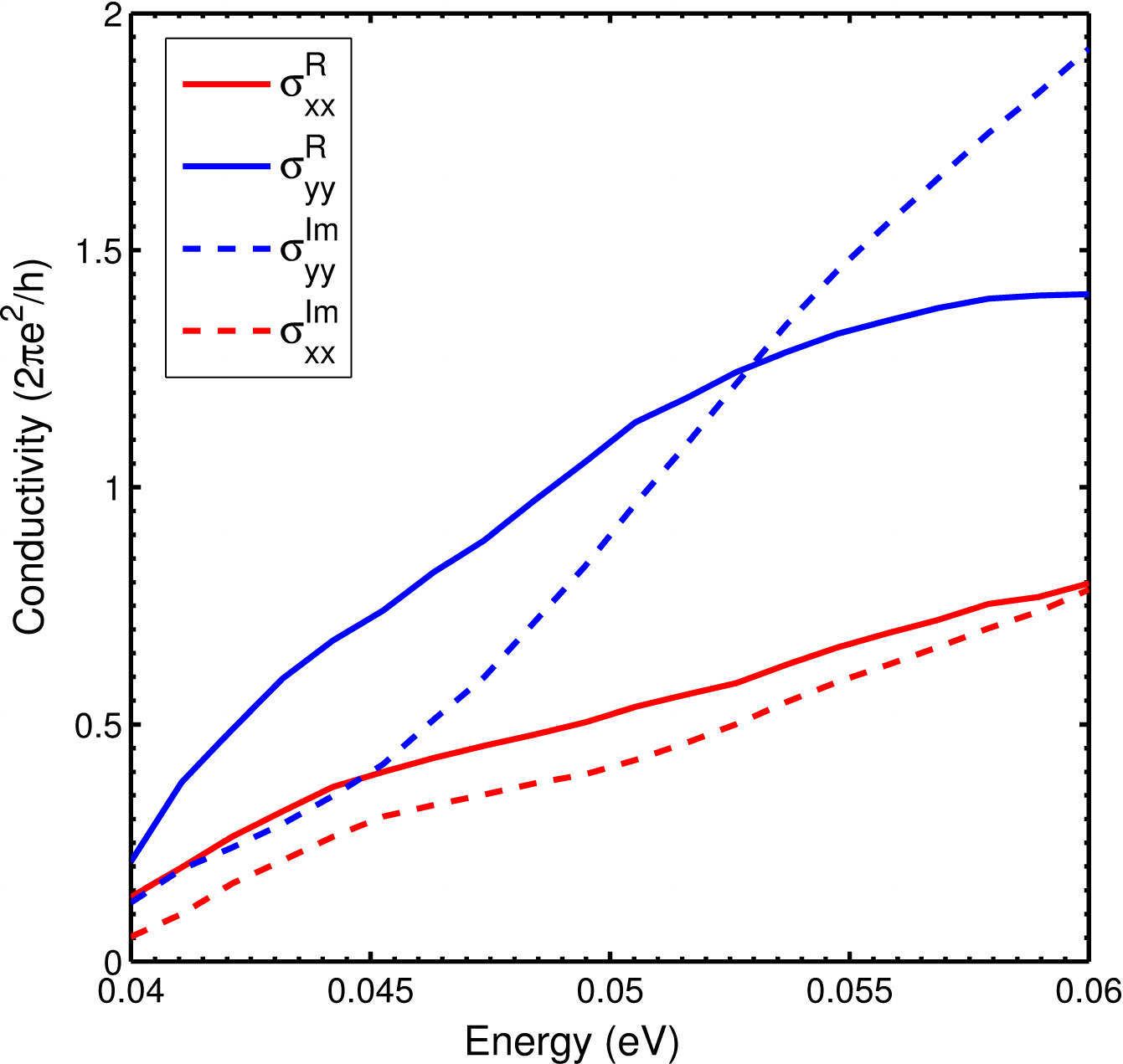}
\caption{The numerically calculated longitudinal optical conductivity of SmB$_{6}$ along the \textit{x}- and \textit{y}-axes. The transit time $\left(\tau\right)$ for surface electrons around the $ \overline{X} $ point on the surface Brillouin zone was taken to be $ 0.1\, ns $. The $ \eta $ in the conductivity equations above is given by $ \eta = \hbar/\tau $. The anisotropy ratio manifested in the unequal surface Fermi velocities is $ v_{f}^{X}/v_{f}^{Y} = 0.7672 $.}
\label{sigmaxxyy}
\end{figure}

The anisotropy (tilted Dirac cone) at the $ \overline{X} $ point suggests that the \textit{y}-axis directed longitudinal conductivity $\left(\sigma_{yy}\right)$ is unequal to $ \sigma_{xx} $. Retracing the set of steps in the calculation of $ \sigma_{xx} $, we write down the result (in units of $ e^{2}/\hbar $) for $ \sigma_{yy} $. Note that the appropriate matrix element in the Kubo expression for $ \sigma_{yy} $ is
\begin{flalign}
\langle\,\Psi_{c}\vert\,\hat{v}_{y}\vert\,\Psi_{v}\rangle  = -\dfrac{iB}{\hbar}\left(\cos\theta + i\gamma\sin\theta\right),
\label{maty}
\end{flalign}
The real part of $ \sigma_{yy} $ is therefore
\begin{align}
\sigma_{yy}^{R} &= \eta\dfrac{B^{2}}{8\pi^{2}}\biggl[\int_{0}^{k_{c}}dk\int_{0}^{2\pi} d\theta\dfrac{\left(\cos^{2}\theta + \gamma^{2}\sin^{2}\theta\right)\Omega}{\left(\hbar\omega - 2\varepsilon\right)^{2} + \eta^{2}} + \nonumber \\
& \int_{0}^{k_{c}}dk\int_{0}^{2\pi} d\theta\dfrac{\left(\cos^{2}\theta + \gamma^{2}\sin^{2}\theta\right)\Omega}{\left(\hbar\omega + 2\varepsilon\right)^{2} + \eta^{2}}\biggr],
\label{tkiry}
\end{align}
while the corresponding imaginary part is 
\begin{align}
\sigma_{yy}^{Im} &= \dfrac{B^{2}}{8\pi^{2}}\biggl[\int_{0}^{k_{c}}dk\int_{0}^{2\pi} d\theta\dfrac{\left(\cos^{2}\theta + \gamma^{2}\sin^{2}\theta\right)\Omega^{'}_{-}}{\left(\hbar\omega - 2\varepsilon\right)^{2} + \eta^{2}} + \nonumber \\
& \int_{0}^{k_{c}}dk\int_{0}^{2\pi} d\theta\dfrac{\left(\cos^{2}\theta + \gamma^{2}\sin^{2}\theta\right)\Omega^{'}_{+}}{\left(\hbar\omega + 2\varepsilon\right)^{2} + \eta^{2}}\biggr].
\label{tkiimy}
\end{align}
For all numerical calculations that are presented, we set the $ \gamma $ term in the conductivity expressions to zero indicating a vanishing band gap at the $ \overline{X} $ point on the surface BZ of SmB$_{6}$. The real and imaginary parts of the anisotropic in-plane longitudinal conductivity is plotted in Fig.~\ref{sigmaxxyy}. We have chosen the energy scale for the conductivity plot to in the range of topological surface states for the slab structure shown in Fig.~\ref{bulksurf}. The inequality of the conductivity components along the \textit{x}- and \textit{y}-axes is clearly visible for the chosen band parameters in the defining Hamiltonian (Eq.~\ref{tkix}. Note that the coefficients $ a_{1} + a_{2} $ and $ a_{1} - a_{2} $ represent the Fermi velocities along the \textit{x-} and \textit{y}-axes, respectively. 

\vspace{0.35cm}
\section{Summary}
\vspace{0.35cm}
In conclusion, we have employed the Kubo formalism from linear response theory to compute the inter-band, and spin conductivity point for a topological insulator with C$_{2v}$ and time reversal symmetry. The Hamiltonian for such a TI to a first order has contributions from the Rashba- and Dresselhaus-like spin-orbit terms. We first show that the longitudinal inter-band conductivities vanishes when the Rashba and Dresselhaus components are of equal strength (the PSH state). At PSH, the zero inter-band longitudinal conductivity is significant since it correlates to a vanishing inter-band light absorption. We also calculated the dynamic longitudinal conductivity of the C$_{2v}$ symmetric surface states of the topological Kondo insulator SmB$_{6}$. The surface states located at the $ \overline{X} $ point of the Brillouin zone host anisotropic Dirac cones; the anisotropy distinguished by unequal Fermi velocities along the \textit{x}- and \textit{y}-axes varies as we progressively move away from the crossing~\cite{roy2014surface}. This asymmetry could be potentially modulated through embedded impurity dopants or inducing strain through a substrate that leads to a controllable optical conductivity with significant implications for production and transmission of chiral surface plasmon polaritons on the 2D anisotropic surface.  

\vspace{0.25cm}
\begin{acknowledgments}
\vspace{0.25cm}
This work at Boston University was supported in part by the BU Photonics Center and U.S. Army Research Laboratory through the Collaborative Research Alliance.
\end{acknowledgments}

%\bibliographystyle{apsrev}
%\bibliography{References}

\begin{appendices}
\appendix
\vspace{0.15cm}
\section{Intra-band optical conductivity}
\label{appA}
\vspace{0.15cm}
To evaluate the intra-band conductivity, we assume that the two energy states $ \varepsilon_{n} $ and $ \varepsilon_{n^{'}} $ are close in magnitude allowing us to Taylor expand the difference between the Fermi functions in Eq.~\ref{kubof}. One then obtains, $ f\left(\varepsilon_{n}\right)-f\left(\varepsilon_{n^{'}}\right) =
\left(\varepsilon_{n} - \varepsilon_{n^{'}}\right)\dfrac{\partial\,f\left(\varepsilon_{n}\right)}{\partial\,\varepsilon_{n}}$. Evaluating the intra-band matrix element $ \mathcal{M}_{intra} = \langle \Psi_{\pm}\vert\,\hat{v_{x}}\vert\,\Psi_{\pm}\rangle $ in Eq.~\ref{kubof} we arrive at the following expression for $ \sigma_{xx}^{intra} $:
\begin{subequations}
\begin{flalign}
\sigma_{xx}^{intra} &= i\dfrac{1}{\mathcal{A}}\dfrac{e^{2}}{\hbar}\sum_{n}\dfrac{\partial\,f\left(\varepsilon_{n}\right)}{\partial\,\varepsilon_{n}}\dfrac{\left(\varepsilon_{n} - \varepsilon_{n^{'}}\right)}{\left(\varepsilon_{n} - \varepsilon_{n^{'}}\right)} \nonumber\\
&\times\, \dfrac{\left(\lambda_{+}\,\lambda_{-}\right)^{2}\left(\alpha_{D}cos\,\theta + \alpha_{R}sin\,\theta\right)^{2}}{\varepsilon_{n} - \varepsilon_{n}^{'} + i\eta}.  
\label{kb2}
\end{flalign}
Using the dispersion relation $ \varepsilon\left(k\right) = \sqrt{\vert\,\beta\,\vert^{2}\,k^{2} + \Delta^{2}}$ to make the change from $ k $ to $ E $ gives
\begin{align}
\sigma_{xx}^{intra} &= \dfrac{e^{2}}{\hbar}\dfrac{1}{4\pi^{2}\eta}\int_{0}^{2\pi}\dfrac{\left(\alpha_{D}f_{1}\left(\phi\right) + \alpha_{R}f_{2}\left(\phi\right)\right)^{2}}{\beta^{2}}\,d\phi \nonumber \\ &\times \int_{0}^{\varepsilon_{f}}\dfrac{\varepsilon^{2} - \Delta^{2}}{\varepsilon}\delta\left(\varepsilon_{f} - \varepsilon\right)\,d\varepsilon,
\label{sumtoint} 
\end{align}
where $ f_{1}\left(\phi\right) $ and $ f_{2}\left(\phi\right) $ are $ \cos\left(\theta\right) $ and $ \sin\left(\theta\right) $, respectively expressed in terms of $ \phi $ using Eq.~\ref{angpo}. Further, in Eq.~\ref{kb2}, we have set $-\dfrac{\partial\,f\left(\varepsilon_{n}\right)}{\partial\,\varepsilon_{n}} = \delta\left(\varepsilon_{n}-\epsilon\right) $ at $ T = 0 $ to rewrite it in the form shown in Eq.~\ref{sumtoint}. The intra-band conductivity normalized to $ e^{2}/\hbar $ is 
\begin{align}
\sigma_{xx}^{intra} = \dfrac{\zeta}{4\pi^{2}\,\eta}\int_{0}^{2\pi}\biggl[\dfrac{\left(\kappa^{2} + 1\right)\cos\phi + 2\kappa\sin\phi}{\left(\kappa^{2} + 2\kappa\,sin2\phi +1\right)}\biggr]^{2}d\phi,
\label{kb6}
\end{align}
\end{subequations}
where $ \kappa = \alpha_{D}/\alpha_{R} $, $ \zeta = \left(\varepsilon_{f}^{2} - \Delta^{2}\right)/\varepsilon_{f} $, and $ \alpha_{R} \neq 0 $. 

\end{appendices}

\end{document}